# What We Read, What We Search:
# Media Attention and Public Attention Among 193 Countries

[Please cite the WWW'18 version of this paper]

Haewoon Kwak
Qatar Computing Research Institute
Hamad Bin Khalifa University
Doha, Qatar
haewoon@acm.org

Jisun An
Qatar Computing Research Institute
Hamad Bin Khalifa University
Doha, Qatar
jisun.an@acm.org

Joni Salminen
Qatar Computing Research Institute
Hamad Bin Khalifa University
Doha, Qatar
jsalminen@hbku.edu.qa

Soon-Gyo Jung
Qatar Computing Research Institute
Hamad Bin Khalifa University
Doha, Qatar
sjung@hbku.edu.qa

Bernard J. Jansen
Qatar Computing Research Institute
Hamad Bin Khalifa University
Doha, Qatar
jjansen@acm.org

## ABSTRACT
We investigate the alignment of international attention of news media organizations within 193 countries with the expressed international interests of the public within those same countries from March 7, 2016 to April 14, 2017. We collect fourteen months of longitudinal data of online news from Unfiltered News and web search volume data from Google Trends and build a multiplex network of media attention and public attention in order to study its structural and dynamic properties. Structurally, the media attention and the public attention are both similar and different depending on the resolution of the analysis. For example, we find that 63.2% of the country-specific media and the public pay attention to different countries, but local attention flow patterns, which are measured by network motifs, are very similar. We also show that there are strong regional similarities with both media and public attention that is only disrupted by significantly major worldwide incidents (e.g., Brexit). Using Granger causality, we show that there are a substantial number of countries where media attention and public attention are dissimilar by topical interest. Our findings show that the media and public attention toward specific countries are often at odds, indicating that the public within these countries may be ignoring their country-specific news outlets and seeking other online sources to address their media needs and desires.

## 1 INTRODUCTION
Foreign news coverage by domestic media outlets shapes public perceptions of foreign countries [49], formulates the public opinion, and offers the basis for policymakers to react to salient issues [42]. The fundamental question, then, is which countries are covered by domestic media outlets and which are not. While there has been a stream of research to reveal systematic factors concerning foreign news coverage [21, 47, 53], the media outlets are not isolated but interact with the audience by monitoring their interests through view counts, likes, or retweets [6, 56]. Therefore, such an approach of focusing only on the media might reveal only a part of the story. However, surprisingly, the spatiotemporal alignment of the news coverage and public interests has been relatively unexplored. As the news industry becomes competitive and the public has easier access to various news sources, the consideration of the alignment with the public interests is required for media outlets to survive.

In this research, we compare and contrast the foreign news coverage and public interests for a wide range of countries. We call them *media attention* and *public attention*, respectively. In order to model the media attention and the public attention, we collect the daily top 100 popular topics for each of 193 countries from Unfiltered News [17], which is an online news aggregator using Google News data, and web search volume data from one country about another country among the same 193 countries from Google Trends [10] during a fourteen-month period from March 7, 2016 to April 14, 2017.

Our research questions are as follows:

- RQ1: How are media attention and public attention structurally aligned?
- RQ2: What is the causal relationship between media attention and public attention?
- RQ3: Are there topical aspects that affect the interaction between media attention and public attention?

By using our longitudinal data collection, we build a multiplex network consisting of the media attention layer and the public attention layer among the 193 countries. The topological analyses of the structures of the multiplex network and time series analyses of link weights on both layers quantitatively address our research questions.

## 2 RELATED WORK
### 2.1 News Coverage as Media Attention
Before the digital era, news outlets were the primary routes for ordinary people to receive information on events taking place abroad. Ever since Galtung and Ruge [8] proposed the idea of news value and its relation to news selection, the studies that followed have revealed various factors [13], such as the power relationship among countries [54, 55], to explain what influences foreign news coverage. Sometimes those factors form a bias in the news presented. For instance, Joye [18] observed a bias in Western news media toward a Euro-American-centered world order. Early studies focusing on a few countries showed inconsistent influential factors due to cultural, regional, and political differences, such as Western society vs. non-Western society [36]. In 1984, Sreberny-Mohammadi conducted a seminal study on foreign news coverage based on a dataset collected from 46 countries [43]. This data has been analyzed further and helped to confirm the critical role of geographical and economic proximity between countries regarding foreign news coverage [44]. New publicly available news datasets, such as the GDELT project [24] and EventRegistry [40], enable researchers to study foreign news coverage on a global scale [3, 4, 21]. Strong regionalism and some power structures among countries have been consistently discovered [18, 21, 47].

## 2.2 Web Search as Public Attention

The conventional way to understand salient issues is to conduct polls; Gallup's "What do you think is the most important problem facing this country today?" has been the de facto standard question to understand salient issues in U.S. society since the 1960s. While such polls are used as primary sources for longitudinal studies of public opinion [16], they require time and resources to scale up.

As Internet connectivity has become pervasive, the online information seeking behavior of the public, such as Discussion in AOL's "Today's News" [38], Yahoo! Buzz Index [14], and Google Trends [28, 50], has been studied as a measure of issue salience. Online information seeking behavior can be collected unobtrusively and efficiently compared to conventional approaches.

## 2.3 Interplay between Media Attention and Public Attention

There have been several attempts to understand the interplay between media attention and public attention. Towers et al. examined the relationship between Internet searches and the news coverage on Ebola in the United States [45]. By using Granger causality, they found that news coverage is a significant driver of web searches. Weeks and Southwell investigated the relationship between Internet searches and the news coverage on the rumor that Barack Obama was secretly Muslim [50], and they discovered the correlation between them peaked on the day of publication. Mukherjee and Jansen [32] studied the effect of advertising during major sporting events. They reported that advertising has a direct impact on web searching. While most of the studies on the relationships between news coverage and Internet searches are based on one or a few topics, our work examines general foreign news coverage and web searches over a fourteen-month period.

## 3 DATA COLLECTION

### 3.1 Unfiltered News as Media Attention

Unfiltered News, which is a web service from Jigsaw (formerly Google Ideas), indexes news articles in the database of Google News and allows users to explore its collection. Particularly, topic- and country-level user interfaces allow users to explore a list of popular topics mentioned by news media in a specific country on a particular day. The topic is an entity in the Google Knowledge Graph [11]. The types of topics, which are the standard schema.org types, such as Person and Place, can be used to distinguish homonyms.

We build our own crawler with a reasonable delay between requests so that the performance of the server is not degraded. We collect the daily top 100 popular topics, which is the maximum number of topics offered by Unfiltered News, in each country for every day from March 7, 2016 to April 14, 2017. For each topic, we collect its metadata using the Google Knowledge Graph API [12]. This metadata contains several fields that are useful for further analyses. These fields are the unique identifier, name, type, and description. Particularly, the unique identifier is used to match the entity between Unfiltered News and Google Trends. We provide detailed explanations of this in Section 3.2. Finally, we collect the topics that are mentioned together with any of the daily top 100 popular topics per country every day, which are called *co-mentions*. For example, if news media outlets in the U.S. mention North Korea due to its nuclear weapons, the co-mentions are the nuclear weapons. As our aim in this work is to examine the attention among countries, we focus only on when the topic concerns another country. As a result, we have 1,322,730 records of media attention between countries and 152,557 unique co-mentions with them.

### 3.2 Google Trends as Public Attention

To collect the data from Google Trends, we use Pytrends [30], which offers a Python interface to access the data that can be accessed via the web interface of Google Trends. The data from Google Trends is a time series of the daily (or weekly) search volume of a given keyword (or multiple keywords) during a given time window. Here, the search volume is not offered as an absolute number but as a normalized value between 0 and 100. The normalization sets the highest search volume in a given time window as 100. The temporal resolution of the search volume is a day when the size of the given time window is small and a week when the size of the time window is big. We find that the time window of seven months is reasonably long and still has the temporal resolution of a day. Thus, we use the time window of seven months in sending a query through PyTrends. Along with the news coverage data collected from Unfiltered News from March 7, 2016 to April 14, 2017, we gather the web search volume data from the same period. That is, for each country, we collect how many Google search queries about other countries are entered during those 404 days by sending two separate requests, each having a time window of seven months. We make an overlap of one day between the two requests. The purpose of that one day is to make a continuous time series of 404 days to have the same normalization scale. We explain in detail how to concatenate two time series data in Section 3.4.

A language issue exists in this process. For instance, the United States is written as 'Los Estados Unidos' in Spanish. This means that, to get the correct search volume from one country about other countries, we need to know their names in the language of the corresponding country. Also, the abbreviation of countries might be another issue. It is not straightforward to find all of the abbreviations for every country in every language. Google Trends addresses this potential issue by using the Google Knowledge Graph Entity

Identifier. This identifier is uniquely assigned to the concept. For instance, an identifier of /m/09c7w0 indicates the concept of the country of the United States of America. Thus, the search volume of /m/09c7w0 contains all the variations of the language and the abbreviations of the United States that Google can recognize. We get the Google Knowledge Graph Entity identifiers of each country through Google Knowledge Graph API [12] and query search volumes by using them.

Like co-mentions in Unfiltered News, Google Trends provides related topics that are the keywords that are searched together for a given keyword. However, with a tight rate limit, it is infeasible to collect all of the related topics. Therefore, in this work, we do not cover the co-search keywords and leave it for future work.

## 3.3 Topic Inference for Co-mentions

The co-mentions collected from Unfiltered News describe the context of media attention from one country to another. We use word embeddings [29] for the topic inference of each co-mention. We use the largest Glove pre-trained word vectors of 840B tokens and 300-dimensional vectors [35]. From manually comparing several news sites, we build a set of 11 common topics, such as world, politics, business, tech, science, health, sports, arts, style, food, and travel. Then, for each co-mention, we compute the distance from each topic and assign the topic to the shortest distance. For example, as the co-mention 'visa' is closest to 'travel' rather than other 10 topics, we regard visa as a travel-related co-mention.

## 3.4 Time Series Concatenation

Figure 1 summarizes the concatenation of two time series from Google Trends. As an example, the figure explains the case when we get the search volume for Korea (KR) and the U.S. (US) from a certain country. The first request is sent with a time window of the seven months from March 7, 2016 to October 7, 2016, and the second request is with a time window from October 7, 2016 to April 14, 2017. To concatenate the corresponding time series from the two requests, we should have a reference that does not change even though Google Trends normalizes the search volumes within a given time window. For that, we use the search volume for the United States on October 7, 2016 (colored as a blue box) as a reference. To handle the potential case when it is zero, we add 1 to the search volumes. As the search volume for the U.S. on October 7, 2016 in the first request becomes 40 by adding 1 to the original search volume (39), to make it to 100, we multiply all the values of the time series in the first request by 100/40=2.5. Similarly, as the search volume for the U.S. on October 7, 2016 in the second request is 80 after the addition of 1, to make it to 100, we multiply all the values of the time series in the second request by 100/80=1.25. Then, as a result, we get the time series from March 7, 2016 to April 14, 2017 by concatenating the corresponding time series from the two requests. By repeating this process, we get a set of the time series representing the search volume from one country to another. Since we use the search volume for the U.S. from one country as a reference in building every time series from the same country, all the resulting time series from that country have the same scale and thus can be comparable.

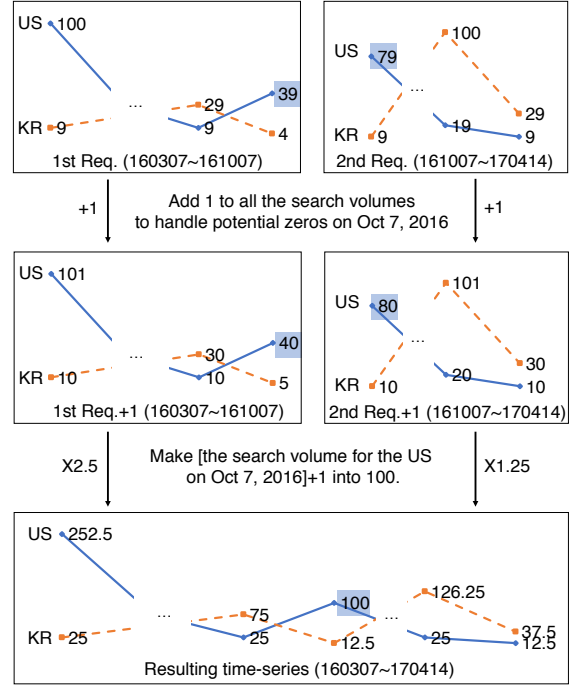

Figure 1: Illustration of time series concatenation

## 4 BUILDING THE MULTIPLEX NETWORK

The multiplex network has multiple layers, and each layer represents a different type of relationship between nodes [20]. For example, phone calls, Facebook posts, or WhatsApp messages can be different layers of a multiplex network representing different types of social interactions. Similarly, we build a multiplex network of attention among countries. The attention has two types: media attention and public attention. The former is captured by news coverage, and the latter is captured by web search volume. Each country is mapped into a node of the multiplex network, and a link is established when the attention from one country to another exists. The link weight represents the strength of the attention.

### 4.1 Media Attention Layer

Media attention is modeled as news coverage of one country by another. If the news media of a country $c_i$ reports on a country $c_j$, we say that media attention exists from the country $c_i$ to the country $c_j$. To capture the temporal dynamics, we do our modeling on a daily basis. On a specific day $t$, a network in the media attention layer (called a media attention network in the rest) denoted by $N_t^M$ has a link from $c_i$ to $c_j$ only when media attention exists from $c_i$ to $c_j$ on day $t$. The weight of links from $c_i$ to $c_j$ is the number of news articles mentioning $c_j$ published by news media in $c_i$. We eliminate self-loops, which represent domestic news.

Once we have $N_t^M$ for all $t \in T$ where T is our data collection period, we build an aggregated network $N^M$ by superimposing $N_t^M$

for all $t \in T$ as follows: 1) if a node or a link exists in any $N_t^M$, it also exists in $N^M$; and 2) the weight of the link from $c_i$ to $c_j$ in $N^M$ is the sum of the weights of the same link in $N_t^M$ where $t \in T$.

The resulting network might have lots of links among countries. Considering the different sizes of the media markets across the countries, it is not trivial to find a global threshold of the link weights for pruning. Therefore, we use a locally computed threshold by a disparity filter [41] with a predefined confidence level to prune the network. Intuitively, this filter finds statistically significant links at every node. The links discovered by the disparity filter are called the *backbone* of the network, which is denoted by $B(\cdot)$.

## 4.2 Public Attention Layer

Public attention is modeled as a web search for one country by people living in another country. If the web search for a country $c_j$ is done by people living in a country $c_i$, we can say that the public attention exists in the country $c_i$ to the country $c_j$. Similar to the modeling of the media attention network, a network in the public attention layer (called a public attention network in the rest) denoted by $N_t^P$ has a link from $c_i$ to $c_j$ only when the public attention exists from $c_i$ to $c_j$ on day $t$. The weight of a link from $c_i$ to $c_j$ on $t$ is the search volume for $c_j$ in $c_i$ on $t$ recomputed in Section 3.4. Also, $N^P$ and $B(N^P)$ comprise the aggregated network of $N_t^P$ for all $t \in T$ and the backbone network of $N^P$, respectively.

## 4.3 Basic Characteristics of the Attention Networks

Table 1 presents the topological characteristics of the media attention network, the public attention network, and their backbone networks, which are the number of nodes $N$, that of links $L$, the mean degree $\langle k \rangle$, the average clustering coefficient $CC$, the assortativity coefficient $\alpha$ [33], the percentage of the strongest connected component $SCC(\%)$, and the reciprocity $\rho$). The superscript$^*$ means that the corresponding measures are computed from the undirected network that consists of reciprocal links of the original network.

|         | N   | L      | $\langle k \rangle$ | CC*  | $\alpha^*$ | SCC  | $\rho$ |
|---------|-----|--------|------|------|-------|------|------|
| $N^M$   | 193 | 23,965 | 124.1 | 0.82 | -0.11 | 1.0  | 0.79 |
| $N^P$   | 193 | 36,864 | 191.0 | 0.99 | -     | 0.99 | 0.99 |
| $B(N^M)$ | 190 | 1,796  | 9.45  | 0.31 | 0.44  | 0.66 | 0.29 |
| $B(N^P)$ | 192 | 1,493  | 7.78  | 0.09 | 0.13  | 0.42 | 0.15 |

Table 1: Basic characteristics of the attention networks

The backbone extraction effectively prunes the networks. Only 7.50% and 4.05% of the links are retained in the backbone of the media attention network and the public attention network, respectively. The assortativity shows an interesting trend. In $N^M$, it is negative; however, in its backbone network $B(N^M)$, it is positive. This means that significant media attention, captured in the backbone network, is exchanged among the countries that have similar degrees because $\alpha > 0$, but such tendency does not appear in general media attention exchange because of $\alpha \approx 0$. As the undirected network converted from $N^P$ consists of one fully connected network and a single isolated node (Rèunion), the assortativity in $N^P$ cannot be computed. The low reciprocity in the backbone reveals the asymmetric nature of the significant media and public attention.

## 5 STRUCTURAL ALIGNMENT BETWEEN MEDIA ATTENTION AND PUBLIC ATTENTION

In this section, we compare the structural characteristics of the media attention and the public attention.

## 5.1 Centrality Correlations

We first focus on important countries in both attention networks. If one country gets major media attention from other countries, does it also get major attention from the public (and vice versa)? To address this question, we compare the centrality of countries in the media attention network and the public attention network. The Spearman correlation coefficients of the degree centrality, betweenness centrality, eigenvector centrality, and closeness centrality between $B(N^M)$ and $B(N^P)$ are 0.562, 0.548, 0.658, and 0.653 (all $p < 0.0001$), respectively. All of these positive coefficients show the moderate positive relationship between the two; the country receiving the high media (public) attention generally gets the high public (media) attention as well, but the alignment is not perfect.

## 5.2 Overlap of Top@k Neighbor Countries

The positive correlations of node centrality imply a structural similarity between media attention and public attention. Then, the question that naturally follows is whether or not such structural similarity can be found at other levels, such as a dyadic level. If a country $c_i$ pays the highest media attention to a country $c_j$, does the highest public attention from $c_i$ also go to $c_j$?

To answer this question in a robust way, we first make the question more general with consideration of the top $k$ countries who get the media (or public) attention from $c_i$. We call these top@k neighbors of $c_i$. For example, top@1 neighbor of one country is the country that gets the most attention from that country, and top@10 neighbors are the top 10 countries that get the most attention from that country. Then, the new question is about whether these top@k neighbors are common or not between $N^M$ and $N^P$.

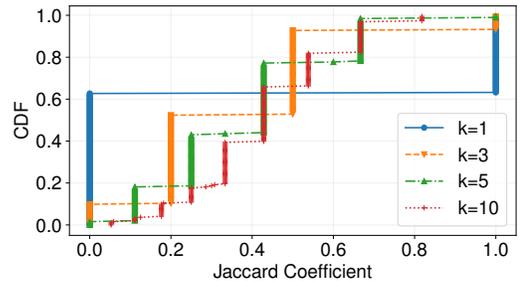

Figure 2: Overlap of top@k neighbors between $N^M$ and $N^P$

Figure 2 shows the cumulative distribution of the Jaccard coefficients of the top@k neighbors for each node between $N^M$ and $N^P$ with varying $k$. The figure shows that the structural characteristics of media attention and public attention are different at the dyadic level, which is opposite to high correlations of node centrality. The media and the public pay the most attention ($k$=1) to different countries in 63.2% of the whole countries, while they pay the greatest attention to the same countries in 36.8% of cases. The median of the Jaccard coefficients is 0.0, 0.2, 0.429, and 0.429 for $k$=1,3,5, and 10, respectively. This shows a structural discrepancy between media attention and public attention at the dyadic level; the media and the public in one country are usually interested in different countries.

Among 122 countries whose top@1 neighbor is different between $N^M$ and $N^P$, the media of 27.0% pay the most attention to the United States, followed by China (8.1%), Syria (5.8%), Egypt (4.1%), and the United Kingdom (5%). These countries have 'news value,' in that they are frequently reported on by the media, but they fail to attract a similar amount of public attention. This can be partially explained by the media's heavy attention to the United States; the media in 64 out of 193 countries pay the most attention to the U.S., but more than a half of them have a public that pays the most attention to other countries instead, such as Germany and Japan. Conversely, in only 10 countries does the public pay the most attention to the U.S. while the media pays attention to other countries.

In addition to the emphasis on the United States by the media, we test two more factors that are known to influence news coverage, which are the distance [47] and trade volumes [53] between countries. Do they also influence public attention? First, we compare the distance from each country to its top@1 neighbor, which is measured by the distance between the capital cities of the two countries, in $N^M$ and $N^P$ by a Mann-Whitney's U test and find no significant difference ($p$-value=0.18,0.16,0.30, and 0.16 for $k$=1,3,5, and 10, respectively). Second, we compile a list of leading export and import trading partners for each country and compare them with top@1 neighbor in $N^M$ and $N^P$. We exclude countries that have 'European Union' as a leading export or import trading partner because it is a union of countries and is not a unit of analysis. As a result, we have a list of 114 countries that have a leading export and import trading partners. Among them, we find that 37 (32.5%) and 34 (29.8%) countries pay the highest attention to their leading importers or exporters in $N^M$ and $N^P$, respectively. No statistically significant difference is confirmed by a Chi-square test ($p$-value=0.7749). Thus, the distance and trade volume between countries do not differently influence the dyadic relationships between $N^M$ and $N^P$.

To investigate whether or not the top@k neighbors are temporally stable, we compute the monthly changes of the median Jaccard coefficients of the top@k neighbors in $N^M$ and $N^P$ with varying $k$. The median Jaccard coefficients are stable, which are 0.0, 0.2, 0.25, and 0.33 when $k$=1,3,5, and 10, respectively, except for the increase in June (0.417) and July (0.429) when $k$=10. By manual inspection, we find that these peaks are driven by two incidents: The United Kingdom European Union membership referendum (Brexit) on June 23, 2016 and the Nice attack resulting in the deaths of 86 people and the injury of 458 others on July 14, 2016. In June 2016, 103 countries paid the highest media and public attention to the United Kingdom, while the United Kingdom's average attention number is 56.3. Also, in July 2016, 125 countries paid the highest media and public attention to France, while its average attention number is only 42.5. The convergence of the media attention in both periods is previously reported in [5]. We additionally observe the convergence of public attention as well as the media attention around Brexit and the Nice attack. Such strong convergence between the media and the public attention is not otherwise observed during the fourteen-month data collection period.

## 5.3 Attention Balance

In the previous subsection, we showed the difference between media and public attention at the dyadic level, particularly focusing on the top@k neighbors, but how much of the attention of one country goes to the top@k neighbors? How skewed are media and public attention from one country to another? To answer this question, we compute the Gini coefficients of the link weights for each node.

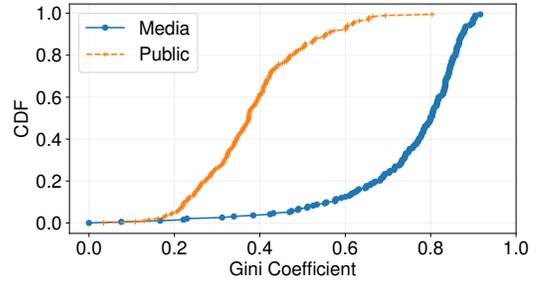

Figure 3: The cumulative distribution of Gini coefficients of the link weights for each node in $N^M$ and $N^P$

The Gini coefficient measures the statistical dispersion or inequality of a given distribution. Basically, a higher coefficient indicates more inequality or a skewed distribution. The minimum value is zero when the given distribution is uniform (perfect equality), and the maximum value is one when the given distribution is maximally skewed. Figure 3 shows the cumulative distribution of Gini coefficients of link weights for each node in $N^M$ and $N^P$. First, the widespread Gini coefficients in $N^M$ and $N^P$ mean that the distribution of link weights is quite diverse across the countries. Some countries, such as Yemen (0.92 in $N^M$) or Gibraltar (0.80 in $N^P$) pay significant attention to a few countries only, but other countries, such as the Central African Republic (0.17 in $N^M$) and Canada (0.13 in $N^P$), pay attention to other countries more equally. Second, the cumulative distribution of the Gini coefficients in $N^P$ is positioned at the left of that in $N^M$. This shows that the public attention goes more equally to other countries than the media attention does. The medians of the Gini coefficients in $N^P$ and $N^M$ are 0.374 and 0.799, respectively. We confirm that the difference of Gini coefficients between $N^P$ and $N^M$ is statistically significant by a Mann-Whitney's U test (U=1887.0, p-value=6.0e-53).

As we expected from Figure 3, the Gini coefficients of each node in $N^M$ and $N^P$ have almost no correlation (Spearman correlation $\rho$=-0.03, $p$-value=0.669). This is additional evidence that the media and the public pay attention to other countries in different ways by focusing on different countries with different strengths.

## 5.4 Attention Flow among Three Countries

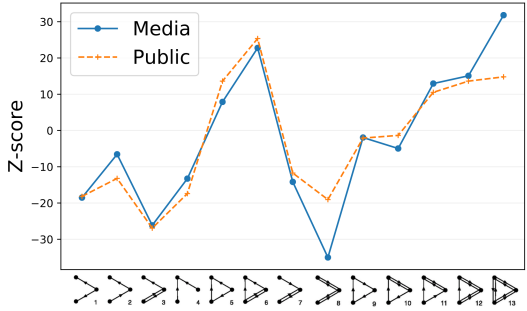

Figure 4: Z-score of 13 motifs in $B(N^M)$ and $B(N^P)$

Through the analysis of top@k neighbors and Gini coefficients of link weights, we can observe the structural differences between media attention and public attention at the dyadic level. We now move on to the more complicated unit of the analysis, a triad sometimes called a three-sized motif. In a directed network, there are 13 different three-sized motifs based on how they are connected. The x-axis in Figure 4 shows a list of them. The intuition behind the motif analysis is that the proportions of each of the 13 different motifs reveal the characteristics of the entire network [31]. For example, a fully connected triad (13th motif in the figure) is overrepresented in a social network than a random network due to the reciprocal nature of human interactions, and in a food web network, a feed-forward loop (FFL, 5th motif in the figure) is overrepresented than in a random network due to the hierarchy in a food chain.

We compute the motif profiles in the backbone of the media attention network and the public opinion network by using Fanmod [51], and the resulting Z-score for each motif is presented in Figure 4. A positive (negative) Z-score means that the corresponding motif is overrepresented (underrepresented) as compared to the null model, which is an ensemble of the 1,000 random networks with the same degree distribution.

We find that both networks show the similar motif profiles; fan-in and fan-out motifs are underrepresented, FFL and double feedback loop motif (6th motif) are overrepresented, and a fully connected motif is overrepresented. Considering that we discovered noticeable structural differences between media attention and public attention in previous experiments, their well-aligned motif profiles are somewhat unexpected. The overrepresented FFL implies the existence of a hierarchy structure in the network. Like a food chain, the FFL shows a transitive hierarchy. Questioning and answering patterns in Yahoo! Q&A also show this hierarchy because a senior expert can answer both easy and hard questions but a junior expert can answer easy questions only [2]. In the media attention and the public attention network, an example of countries forming the FFL is [Algeria → France, Algeria → the U.S., and France → the U.S.]. The public in Algeria pays significant attention to France and the U.S. and that in France to the U.S., but the public in the U.S. does not pay significant attention to either of them. Likewise, influential countries (e.g., G8), such as the United Kingdom [Luxembourg → Belgium, Luxembourg → the United Kingdom, Belgium → the United Kingdom] or Japan [Malaysia → Bangladesh, Malaysia → Japan, Bangladesh → Japan], form many FFLs in both attention networks. Many of the overrepresented fully connected motifs are observed from geographically closely located countries, such as [Brazil, Argentina, Paraguay], [Croatia, Bosnia and Herzegovina, Serbia], and [China, Hong Kong, Macau]. As neighboring countries are likely to share linguistic, historical, or cultural similarities, it is reasonable that their media and public pay attention to each other.

## 5.5 Attention Flow among Regions

Here we add geographical information as the node properties to the network to extract meaningful patterns from the attention flow. While a strong regionalism has been found in news coverage [21], recent studies also report the 'global village' trend where several tens of countries exchange media attention beyond regional blocks [22]. Then, which of these views more accurately describes the media attention and the public attention captured in our data collection? By comparing how different regions exchange media attention and public attention, we aim to gain insight into how the attention flows in the world.

|          | Africa | Americas | Asia | Europe | Oceania |
|---------:|:------:|:--------:|:----:|:------:|:-------:|
| Africa   | 0.19   | 0.05     | 0.63 | 0.13   | 0.00    |
| Americas | 0.05   | 0.32     | 0.37 | 0.25   | 0.01    |
| Asia     | 0.12   | 0.11     | 0.60 | 0.16   | 0.01    |
| Europe   | 0.04   | 0.21     | 0.31 | 0.42   | 0.01    |
| Oceania  | 0.05   | 0.28     | 0.34 | 0.24   | 0.10    |

(a) Media Attention

|          | Africa | Americas | Asia | Europe | Oceania |
|---------:|:------:|:--------:|:----:|:------:|:-------:|
| Africa   | 0.31   | 0.19     | 0.23 | 0.24   | 0.03    |
| Americas | 0.15   | 0.41     | 0.20 | 0.21   | 0.04    |
| Asia     | 0.15   | 0.17     | 0.42 | 0.21   | 0.04    |
| Europe   | 0.13   | 0.18     | 0.21 | 0.43   | 0.04    |
| Oceania  | 0.13   | 0.20     | 0.28 | 0.17   | 0.22    |

(b) Public Attention

Figure 5: Attention flows among regions in $N^M$ and $N^P$. The row is a source region, and the column is a destination region. Flows are normalized within each source region.

Figure 5 shows the normalized attention flow from the source region (row) to a destination region (column). We normalize the sum of the row to 1. We find some similarities and differences between media attention and public attention as shown in the table. First, a strong regionalism can be found in both but with a slightly different magnitude. Regarding media attention, Asia and Europe pay attention to themselves the most, but other regions do the second most. By contrast, regarding public attention, except Oceania, all the regions pay the most attention to themselves. Second, the emergence of Asia as a receiver of the media attention is noticeable, while it is not visible in the public attention. All of the regions except for Europe pay the highest level of attention to Asia, and even Europe pays the second-highest level. Finally, the public pays attention more equally than the media does. This is what we found at the country level in Section 5.3, and it is again observed at the region level. The median Gini coefficients of the media and the public are 0.439 (Europe) and 0.317 (Americas). We confirm that the difference is statistically significant by Mann-Whitney's U test (U=3.0, $p$-value=0.030).

## 5.6 Community Structure in Networks

In the previous section, we use a geographical boundary as a notion of a group of countries and look into attention within and across the groups. Now we find groups of countries based on the actual attention flow by a community identification method. The resulting groups can reveal an embedded modular structure of the media attention and the public attention networks.

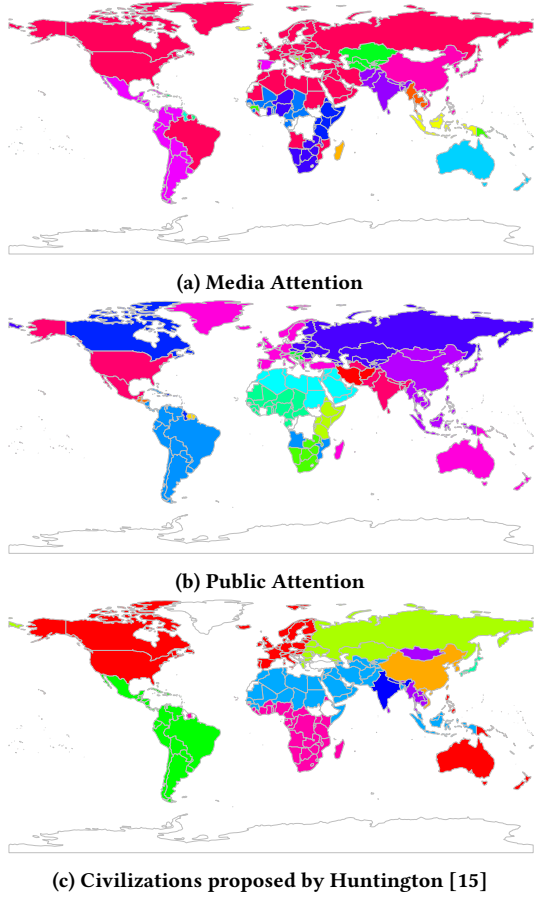

(a) Media Attention

(b) Public Attention

(c) Civilizations proposed by Huntington [15]

Figure 6: Community structure in backbone networks

Among various community identification methods, we use InfoMAP [39] based on a random walk that can incorporate the direction of the links by nature. We visualize the identified communities in $B(N^M)$ and $B(N^P)$ in Figure 6(a) and (b), respectively. The countries in the same community are painted the same color. We find 16 and 14 communities in $B(N^M)$ and $B(N^P)$, respectively. From the figure, the primary difference is the coverage of the biggest community. In $B(N^M)$, there is one giant community of 80 countries from the whole world, such as Europe, the Middle East, North America, and Africa, showing the 'global village' trend of media attention previously reported in [22]. While the membership of the countries in the biggest community is slightly different from [22], the pattern that the media attention does not stay within a geographical region but is exchanged globally is consistent. By contrast, the biggest community in $B(N^P)$ has 42 countries, which mainly consist of European and Oceanian countries. No country in other regions is included in the biggest community in $B(N^P)$. Instead, regional splits are clear where public attention is exchanged.

Interestingly, several groups of geographically closely located countries bring the notion of civilizations proposed by Huntington [15] to mind. He argued that "civilizations" sharing a cultural and religious identity become important to understand conflicts in the post-Cold War world. His concept of civilization has been widely used to understand the division of the world, for example, in [34]. Figure 6(c) shows the map of civilizations. We can see that some of the civilizations appear as the identified communities in $B(N^M)$ and $B(N^P)$ with the addition or removal of few countries. Revealing this fine-grained variation from the civilizations originally proposed in the late 90s [15] poses an interesting research challenge. Are the cultural lines Huntington discovered changing due to connectivity and globalization trends? The fact that the core of civilizations is still consistently observed in media attention and public attention proves the validity of the original civilizations, but the trend of the global village and other variations require new models and explanations for today's international relations.

# 6 INTERPLAY BETWEEN MEDIA ATTENTION AND PUBLIC ATTENTION

We have shown structural similarities and differences between media attention and public attention at different levels. We then move on to the interplay between these attentions.

## 6.1 Granger Causality Analysis

Granger causality is a statistical concept of causality between two time series. One time series $T_X$ *Granger-causes* the other time series $T_Y$ when past values of $T_X$ can improve the explanation of the current value of $T_Y$ compared to when past values of $T_Y$ are used alone. Granger causality has been widely used for causation analysis due to its computational simplicity. Our aim is to examine whether media attention Granger-causes public attention, public attention Granger-causes media attention, or both. To achieve this, for each country pair that exists both in $N^M$ and $N^P$, we build two time series representing media attention and public attention each. We denote by $T_M^{c_i \to c_j}$ the time series of the media attention from a country $c_i$ to a country $c_j$, and $T_P^{c_i \to c_j}$ the time series of the public attention from $c_i$ to $c_j$.

To examine whether media attention Granger-causes public attention, the null hypothesis is that $T_M^{c_i \to c_j}$ does not Granger-cause $T_P^{c_i \to c_j}$. We test the hypothesis with varying time lags from 1 to 14 days. We reject the hypothesis when $p$-value < 0.05 and say that media attention Granger-causes public attention with a time lag $l$. If two time series show a Granger-cause relationship with multiple $l$s, we choose the $l$ with the largest $F$ statistic (the smallest $p$-value). For the case that public attention Granger-causes media attention, the null hypothesis is that $T_P^{c_i \to c_j}$ does not Granger-cause $T_M^{c_i \to c_j}$, and the rest of the process is the same.

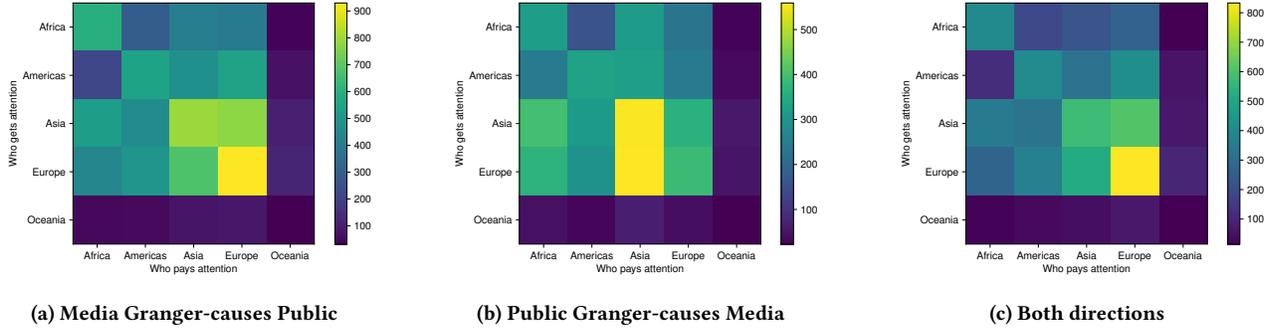

(a) Media Granger-causes Public  (b) Public Granger-causes Media  (c) Both directions

Figure 7: Granger-causal relationships among regions

For the hypothesis testing, we note that we make time series stationary through appropriate transformation techniques such as differencing. Among 37,056 country pairs, we find that media attention Granger-causes public attention in 2,594 country pairs, and public attention Granger-causes media attention in 5,817 country pairs. Also, in 6,698 country pairs, we find that Granger-causal relationships in both directions exist. We illustrate how those relationships are geographically distributed in Figure 7.

Figure 7(a) shows the country pairs where media attention Granger-causes public attention, (b) shows the country pairs where public attention Granger-causes media attention, and (c) shows the country pairs who have both Granger-causal relationships (an intersection of (a) and (b)). By comparing Figure 7 with the attention flows among regions in Figure 5, we see some similarities and differences. In Figure 7(c), we can see that a stronger association between media attention and public attention when the attention pays to a country in the same region than to the other regions. The media attention or the public attention within the same region is not only higher than that across the regions (Figure 5) but also is likely to Granger-cause to its counterpart. However, the higher volume of media (or public) attention from one region to another does not guarantee that it Granger-causes more public (or media) attention in the corresponding pair of regions. This disagreement between media attention and public attention is consistent with a previous study comparing news articles and news comments [1].

We then investigate what determines the Granger-casual relationships. For this, we build a random forest model to predict whether a relationship from a source country to a destination country is divided into four classes: (1) media attention Granger-causes public attention, (2) public attention Granger-causes media attention, (3) both Granger-casual relationships exist, or (4) no Granger-casual relationship exists. For both source and destination countries, we incorporate tens of country properties, such as population, GDP, literacy rate, Internet penetration, press freedom index, and so on. Also, we consider topological characteristics of countries in $N^M$ and $N^P$. Furthermore, we consider the distance between two countries for modeling the stronger 'same region' attention. As the number of the relationships for each class is largely different, we down-sample the relationships for handling a class imbalance issue and make the number of observations for each class the same. We split the data 75% and 25% for training and testing, respectively. The accuracy of our resulting model is 99.81% better than the random guess baseline. We note that the goal of this model is to find the important variables for causal relationships rather than to build the best prediction model.

The top five important variables from the trained model show what influences on Granger-causal relationships. They are: distance between two countries, out-degree in $B(N^M)$, GDP of the source country, the Internet penetration in the source country, and the GDP of the destination country. This is consistent with previous work that reveals the crucial role of the distance between countries and the GDP of a destination country for the news coverage by a source country [21, 53]. Also, other variables that can affect the behavior of information gaining and seeking can influence media attention and public attention.

## 6.2 Topic-level Interaction

As we mentioned earlier, co-mentions reveal the context of media attention. Prior to the analysis of the topic-level interaction between the media attention and the public attention, we examine why each country is covered by the rest of the world based on co-mentions.

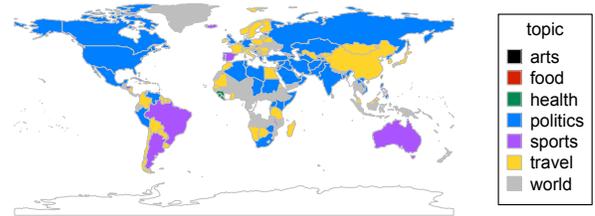

Figure 8: Why is each country covered by other countries?

Figure 8 shows topics of co-mentions for each country. Of course, a country is covered due to a variety of reasons. To address this, for country $c_i$, we first check how many times each topic is co-mentioned by $c_j$. Then, we can find that the topic with the most appearance is the topic of media attention from $c_j$ to $c_i$. By collecting the topics of media attention from each country to $c_i$ and choosing the topics featured in most of the countries, we have the topic of why $c_i$ is covered by the rest of the world.

The association between a country and its topic fits in with our intuition well. Among the G8 countries, five (Germany, the U.K., the U.S., Canada, and Russia) are covered with 'politics,' and three (France, Italy, and Japan) are covered with 'travel.' France and Italy are actually the most popular and the fifth-most popular countries by international tourist arrivals, respectively, and Japan showed double-digit growth five years in a row [46]. Also, popular destinations in Africa are listed by one of the famous travel guides, Lonely Planet. These include the Giza Pyramids in Egypt, the Okavango Delta in Botswana, and Etosha National Park in Tanzania, are also covered with 'travel.' In contrast to negative stereotyping of the third world by media attention in the early 2000s [9], the emergence of balanced news coverage signifies the advancement of the media industry, while most of the other African countries are still covered with 'politics.' Countries that are known for football, such as Brazil, Argentina, or Spain, are covered with 'sports.'

We also examine how one country is covered by a set of countries in a specific region instead of the rest of the world. Due to the lack of the space, we omit the visualizations but share some common trends. First, the United States, the United Kingdom, and Russia are always covered with 'politics' by any region, showing their influence on today's world. Second, countries get covered with 'travel' more by the same region than other regions possibly due to its accessibility. Finally, we see several consistent connections between a country and a topic across the regions, such as Brazil and sports, and most of the Middle Eastern countries and politics.

We repeat the same Granger causality analysis as we did in Section 6.1 by using topical media attention and public attention. While the overall trend is similar as before, such as a strong association between media attention and public attention within the same region, there are also interesting topical trends. The top five topics that lead to the most Granger-causal relationships in both directions are sports, travel, world, politics, and arts. High engagement of the public in the entertainment-oriented news (often called soft news [37]) and supplies of such news by the media are well captured here. In the trained model, we achieve similar performance, which is 91.03% higher than the random baseline. The topic becomes the most important variable, as we expected, in addition to the top five variables in an earlier model.

## 7 DISCUSSION

In this work, we examined the media attention and the public attention among 193 countries by using the data from Unfiltered News and Google Trends during more than one year. We discovered that the structural characteristics of media attention and public attention are similar at some levels and different at others: (1) the importance of a node (centrality) is positively correlated between two networks; (2) to whom one country pays significant attention is different between two networks; (3) the distribution of attention across neighbors is different between two networks, public attention being distributed more equally than the media attention; (4) the relationship among three countries is similar between two networks, with overrepresented feed-forward loops implying the transitive hierarchy of the attention, and the overrepresented fully connected triads meaning that attention is exchanged among geographically closely located countries, and (5) community structures of media attention show the trend of a global village, while those of public attention show clearer geographical splits. These similar and different two networks interact with each other in terms of Granger causality. We showed that media attention and the public attention within the same region strongly associate with each other. Along with distance, country properties influence the Granger causality. There are some variations in the interplay between the media attention and the public attention according to the topics.

The structural alignment of media attention and public attention shows the interaction between the news media and the public opinion [27]. On the one hand, the agenda-setting effects of the media influence public perceptions on foreign countries [49], build consensus [25], and thus help policymakers understand the public issue salience [42]. On the other hand, as social media becomes one of the most important channels with which to disseminate news [23], it becomes easier for the media to collect public opinion about an issue. Also, reader metrics, such as clickstreams, likes, or shares, exert pressures on news selection by the media [6, 48]. Despite the interaction between the media and the public, the existence of the structural difference between the media attention and the public attention makes us revisit the old problem about the role of the journalists, editors, and media. It has long been debated where the line should be between what people want to read and what people ought to know [52]. Also, the transfer from media salience to public salience does not always necessarily happen [19]. For example, when the news is not relevant for the audience or is clear without any uncertainty, the transfer is less likely to occur [26].

Across our experiments, two trends are consistently found: the global village trend and strong regionalism. The global village trend is more noticeable in the media attention than the public attention, which has been reported as the media convergence [3, 7]. The international news agencies, such as AP, AFP, and Reuters, play an essential role in this [21, 53]. The clearer geographical splits in the public attention imply that the cultural affinity within a region matters more to individuals' information-seeking behavior.

There are potential limitations in this work. First, we model media attention and public attention from a single service each, which are Unfiltered News and Google Trends, respectively. However, since their actual data is aggregated from tens of thousands of news media and billions of individuals, the problem of generalizability is relatively limited. Second, we set the scope of the analysis as the foreign news coverage and the corresponding web search. This makes mapping the news coverage and the web search by using the names of countries as the unit of the analysis straightforward; nevertheless, at the same time, this does not cover the domestic media and public attention. Considering the aim of this research is to understand the attention among countries that can influence public perception of foreign countries and foreign policy, the domestic news and search do not distort the result. However, incorporating them as well would be an interesting direction for future research.

## ACKNOWLEDGMENTS
The authors would like to thank Prof. Yamir Moreno from University of Zaragoza and Prof. Michael W. Macy from Cornell University for their valuable comments in early phases of the research.